\documentclass[conference]{IEEEtran}
\IEEEoverridecommandlockouts

\usepackage{booktabs} 
\usepackage[table]{xcolor}
\usepackage{multirow}
\usepackage{tablefootnote}
\usepackage[flushleft]{threeparttable}
\usepackage{amsmath}
\usepackage{enumitem}
\usepackage{cite}
\usepackage{amsmath,amssymb,amsfonts}
\usepackage{algorithmic}
\usepackage{graphicx}
\usepackage{textcomp}
\usepackage[utf8]{inputenc}

\def\BibTeX{{\rm B\kern-.05em{\sc i\kern-.025em b}\kern-.08em
    T\kern-.1667em\lower.7ex\hbox{E}\kern-.125emX}}

\begin{document}

\title{Activity-Based Analysis of Open Source Software Contributors: Roles and Dynamics}

\author{
\IEEEauthorblockN{Jinghui Cheng}
\IEEEauthorblockA{\textit{Department of Computer and Software Engineering} \\
\textit{Polytechnique Montreal, Montreal, Canada}\\
jinghui.cheng@polymtl.ca}
\and
\IEEEauthorblockN{Jin L.C. Guo}
\IEEEauthorblockA{\textit{School of Computer Science} \\
\textit{McGill University, Montreal, Canada}\\
jguo@cs.mcgill.ca}
}
\maketitle
\begin{abstract}
Contributors to open source software (OSS) communities assume diverse roles to take different responsibilities. One major limitation of the current OSS tools and platforms is that they provide a uniform user interface regardless of the activities performed by the various types of contributors. This paper serves as a non-trivial first step towards resolving this challenge by demonstrating a methodology and establishing knowledge to understand how the contributors' roles and their dynamics, reflected in the activities contributors perform, are exhibited in OSS communities. Based on an analysis of user action data from 29 GitHub projects, we extracted six activities that distinguished four Active roles and five Supporting roles of OSS contributors, as well as patterns in role changes. Through the lens of the Activity Theory, these findings provided rich design guidelines for OSS tools to support diverse contributor roles.
\end{abstract}


    
    

\section{Introduction}
\label{sec:Introduction}
As a software development model, OSS has experienced a fast growth during the past decades. The communities around OSS projects are becoming increasingly heterogeneous, comprising not only developers and tech-savvies but also designers, managers, and users with a wide-ranging level of experience and expertise. As a result, the ways participants contribute to the OSS projects also become increasingly diverse \cite{Cheng2018, GitHub2018}. However, one major limitation of the current OSS tools and platforms is that they provide a uniform user interface regardless of the activities performed by the various types of contributors interacting with the platform. In other words, the current OSS tools do not take into enough account the various roles assumed by the OSS contributors.

This paper serves as a non-trivial first step towards resolving this challenge by demonstrating a methodology and establishing knowledge to understand how the roles and their dynamics are currently exhibited in OSS communities. In particular, we focused on examining the roles and their dynamics based on the types of \textit{activities} that OSS contributors perform. This perspective is inspired by several aspects of the Activity Theory \cite{Vygotsky1981}. Particularly, the Activity Theory, applied to the field of Human-Computer Interaction, specifies that information and communication tools need to focus on mediating human activities, facilitating users to perform a group of low-level actions and operations in order to achieve higher-level objectives; additionally, such mediation needs to be adjustable in an evolving context \cite{Nardi1996}. To guide our study, we pose the following research questions:
\begin{enumerate}[leftmargin=1cm,label=\textbf{RQ{{\arabic*}}}:]
\item What are the decisive activities that distinguish the roles assumed by OSS contributors?
\item What are the prominent roles that can be identified through analyzing a wide range of actions community contributors perform in a diverse set of projects?
\item How do the roles assumed by the community contributors in an OSS project change over time?
\end{enumerate}
To answer these questions, we collected and analyzed action data of 20,838 unique contributors from 29 diverse GitHub projects and conducted factor and clustering analyses to identify the prominent activities and roles of OSS contributors. 

In the following sections, we first briefly review the related work (Section \ref{sec:RelatedWork}). We then outline our data collection and analysis methods (Section \ref{sec:Methods}). In Section \ref{sec:ResultsAndDiscussion}, we report our results on the identified activities and roles, as well as the role dynamic patterns. We then discuss the implications of our findings to the design of OSS tools (Section \ref{sec:Implications}). Finally, we provide concluding remarks in Section \ref{sec:Conclusion}
\section{Related Work}

\label{sec:RelatedWork}
Our study is related to previous work that focused on users' roles in information and communication technologies (ICTs) and studies that investigated the structure of OSS communities.

\subsection{Role-Based Approaches in ICTs}
Previous work has explored models and techniques to identify and support different roles in ICTs of various application domains, including collaboration tools \cite{Barlow2013}, access control systems \cite{BenFadhel2018}, knowledge co-production platforms \cite{Arazy2016}, and software engineering tools \cite{Zhu2006, Acuna2004}. For example, Arazy et al. \cite{Arazy2016} identified seven roles of Wikipedia contributors, such as \textit{all-round contributors} and \textit{layout shapers}, through a clustering analysis of user actions in Wikipedia articles.

Our study is most closely related to previous work that investigated roles in tools and techniques that support software design and development \cite{Tomakyo2005}. Zhu et al. \cite{Zhu2006} advocated a complete and consistent role consideration in all aspects of software engineering and in research about tools through the lens of Role-Based Software Development (RBSD). 
Acuña and Juristo \cite{Acuna2004} also proposed a model that consists of 20 general capabilities crucial in software development and mapped these capabilities with 20 predetermined roles in software projects. Leveraging this model, they presented a procedure for assigning people to roles according to their capabilities.

More recently, researchers investigated role dynamics in self-organized software development teams. Hoda et al. \cite{Hoda2013} conducted Grounded Theory research involving 58 agile practitioners from 23 software organizations to understand the role dynamics in agile teams. They identified six ``informal, implicit, transient, and spontaneous'' roles performed by practitioners to reinforce the self-organizing nature of agile practice. These roles include, for example, \textit{mentors} who guide and inform the team in using agile methods, \textit{translators} who communicate between customers and technical team, and \textit{champions} who acquire supports from senior management. Our study builds upon these work and explores the dynamics of various activity-based roles in OSS communities.

\subsection{OSS Community Structure}
Much work that investigated the OSS community structure is based on the ``onion'' model \cite{Nakakoji2002, Ye2003}. This model proposed a layered structure of responsibilities for OSS projects that included a small number of core members and a larger number of peripheral developers and bug fixers \cite{Nakakoji2002}. Mockus et al. \cite{Mockus2002} examined the Apache web server and the Mozilla browser as case studies and empirically generated several hypotheses concerning the OSS community structure. These hypotheses echoed with the ``onion'' model that a small number of developers 
contributes to the majority of the codebase.

Because of the self-organizing nature of OSS communities \cite{Hoda2013}, researchers have particularly investigated the evolution of the OSS structure \cite{Onoue2014, Cheng2017, Joblin2017a}.
For example, Cheng et al. \cite{Cheng2017} identified several factors that significantly influenced developers’ evolution into a core member in OSS ecosystems; such factors included the total number of projects developer were willing to join and the degree to which the developer's peers were closely connected. Joblin et al. \cite{Joblin2017a} also identified that the OSS communities tended to evolve from a hierarchical structure to a hybrid one with a greater distribution of contributions while the number of developers increases.

More closely related to our work, several recent studies focused on exploring classification methods for OSS communities. Through a clustering analysis on code committing metrics extracted from ten OSS projects, Di Bella et al. (2013) identified three major factors and four developer role groups that fell on the spectrum from core to occasional rare developers \cite{DiBella2013}. Agrawal et al. (2016) also adopted a clustering approach and explored decision tree models to classify OSS code committers; their developer classes also ranged from core developers to less engaged developers \cite{Agrawal2016}.

A major limitation of these studies is that they only focused on code committing activities. While we adopted a similar statistical approach, our study focused on a much wider variety of actions beyond code contribution and identified more descriptive activity-oriented factors and roles. We also aimed to extract common roles in a wide range of OSS projects.

\vspace{10pt}
In sum, while previous work has demonstrated non-negligible effort on understanding the roles and their structures in ICTs and OSS communities, there are seldom explicit investigations, with the aim of improving tool support in OSS, on the correlations among actions performed during goal-driven activities, nor on the dynamics of activity migrations accompanied by frequent role changes. Therefore, our work fills the gap by exploring those important aspects with a data-driven approach and by following up with a detailed discussion on the implications for OSS tool design. 

\section{Methods}
\label{sec:Methods}
We analyzed user action data within the last three years from 29 GitHub projects that exhibit diverse characteristics. All data was collected in January 2018.

\subsection{Projects Selection}
\label{subsec:Methods-Projects}
\begin{table*}
\centering
\caption{Names of GitHub projects selected to our study}
\label{Table:ProjectNames}
\small
\begin{tabular}{llllll}
\toprule
accessibility-developer-tools & better\_errors & hospitalrun-frontend & neovim & refined-github & the\_silver\_searcher \\
adarkroom & brew & jekyll & picongpu & SoundManager2 & TrueCraft \\
advocacy.mozilla.org & cocos2d-html5 & kubernetes & primer & spine & urh \\
artsy.github.io & csslint & madison & pysc2 & superpowers-core & utron \\
basscss & guardian/frontend & mention-bot & railsbridge/docs & swipl-devel \\
\bottomrule
\end{tabular}
\end{table*}
To cover a wide range of OSS communities, we focused on projects in different application domains. Particularly, we randomly selected one project in each category in GitHub ``Collections'' \footnote{https://github.com/collections/}. GitHub ``Collections'' are curated lists (a total of 31 lists at the time of our data collection) of recently active and influential projects and communities. We eliminated two lists, ``Open data'' and ``Policies'', which focused on non-software projects. Table \ref{Table:ProjectNames} includes the names of the selected projects. These projects involved a total of 20,838 unique contributors (including code contributors, issue reporters and discussion participants, and pull request reporters and discussion participants), 41,275 issues, 73,763 pull requests, and 240,024 commits. The code repositories are comprised of 4,963,540 lines of code in 24,451 files, covering 15 programming languages.

\subsection{Metrics Selection}
\label{subsec:Methods-Metrics}
To effectively assess the participants' contribution to their OSS community, we selected metrics gathered from various aspects. Those metrics describe the detailed \textbf{\textit{actions}} contributors take in order to participate on the OSS projects. First, \textit{code contribution} metrics include numbers of commits made, lines of code changed, and files edited, as well as metrics related to pull requests (PRs) made by contributors. Second, \textit{opinion contribution} metrics assess actions associated with reporting issues and commenting in issue and PR discussions. Third, \textit{network-related} metrics include the number of times a participant was mentioned or referred other issues or PRs in discussions. Finally, \textit{administration} metrics measure managerial actions such as managing labels or manipulating issues or PRs. Those metrics were inspired by several previous works \cite{DiBella2013, Agrawal2016, Joblin2017} and are summarized in Table \ref{Table:Metrics}.

\begin{table}
\renewcommand{\arraystretch}{0.8}

\begin{threeparttable}

\centering
\small
\caption{Action metrics to assess OSS participant's contribution}
\vspace{-5pt}
\label{Table:Metrics}
\begin{tabular}{@{}cl@{}}

\toprule
\textbf{Type} & \multicolumn{1}{c}{\textbf{Metric}} \\

\midrule
\multirow{5}{*}{\begin{tabular}[c]{@{}c@{}}Code\\Contrib.\end{tabular}} & \# of commits made \\
 & \# of line of code changed in the codebase \\
 & \# of files worked on \\
 & \# of pull requests (PRs) made \\
 & Avg. length of PR descriptions* \\

\midrule
\multirow{6}{*}{\begin{tabular}[c]{@{}c@{}}Opinion\\Contrib.\end{tabular}} & \# of issues reported \\
 & Avg. length of issue descriptions* \\
 & \# of comments made in issue discussions \\
 & Avg. length of issue comments* \\
 & \# of comments made in PR discussions \\
 & Avg. length of PR comments* \\

\midrule
\multirow{4}{*}{Network} & \# of times being mentioned in issue comments \\
 & \# of times being mentioned in PR comments \\
 & \# of times referred other issues/PRs in issue comments \\
 & \# of times referred other issues/PRs in PR comments \\

\midrule
\multirow{4}{*}{Admin.} & \# of times applied or removed labels on issues \\
 & \# of times applied or removed labels on PRs \\
 & \# of times closed issues \\
 & \# of times closed pull requests\\

\bottomrule
\end{tabular}
\begin{tablenotes}
\item* All lengths were measured in number of characters
\end{tablenotes}
\end{threeparttable}
\vspace{-10pt}
\end{table}

\subsection{Data Collection}
We aimed at extracting the necessary metrics from the repositories of the 29 GitHub projects and focused on the contributor actions within the three-year period between January 1st, 2015 and January 1st, 2018. To collect such a data set, we first used the GitHub REST API \footnote{https://developer.github.com/v3/} to download the raw data about code committing actions, issue reporting and commenting actions, PR reporting and commenting actions, as well as issue and PR events (e.g. labels applied/removed, closed, etc.) for each project. We then excluded any action data performed by ``bots'' (i.e. automated processes presented as GitHub users who perform event-driven actions). In order to understand the dynamics of the OSS roles, data for each contributor was then divided based on the quarter of a year when we calculated the metrics. As such, our data set accumulated metrics for each participant in each project across 12 time periods. In total, this data set is comprised of 38,891 data points, each included 19 dimensions corresponding to the metrics described in Table \ref{Table:Metrics}.

\subsection{Identifying Activities and Roles}
\label{subsec:factor_cluster}
The metrics introduced previously in Section \ref{subsec:Methods-Metrics} were selected to measure the concrete actions taken by the user from distinct perspectives. Those metrics, however, might be interrelated and can be influenced or determined by a set of hidden factors. We hypothesize that those hidden factors are the common \textbf{\textit{activities}} that OSS contributors engage in when they are serving certain roles in the projects. To identify these activities, we first performed a Factor Analysis on the dataset to understand and interpret the interrelations between those metrics. Based on these factors, we then conducted a Clustering Analysis to identify the prominent contributor roles. Before the factor analysis, all metrics were standardized to have a mean of zero and unit variance.

\subsubsection{Factor Analysis}  
\label{subsubsec:factor}
Factor analysis, especially exploratory factor analysis, is a statistical method to discover underlying patterns in a set of variables \cite{child2006essentials}. The main procedures for factor analysis include factor extraction and rotation.

Maximum Likelihood and Principal Axis Factors (PAF) are two commonly adopted factor extraction techniques \cite{costello2005best}. We chose the PAF approach because preliminary analysis indicated that the distributions of our data violate the assumption of multivariate normality \cite{fabrigar1999evaluating}. After extracting the factors, we used the Kaiser criterion and retained the factors with an eigenvalue larger than 1.0, indicating that those are the most influential factors (i.e. factors that account for the most variance in the data) \cite{zwick1986comparison}.

The retained factors were then rotated to attain a simple structure that supports a better interpretation. In such a structure, each rotated factor aims to define a distinct group of interrelated metrics. Rotation techniques can be generally divided into orthogonal and oblique rotations; the former produces factors that are uncorrelated while the latter allows the factors to correlate. In social science, behaviors can rarely be partitioned into groups that are independent \cite{costello2005best}. We hypothesized that the factors influencing the contributors' activities in OSS communities would also exhibit some correlations. We therefore decided to use the oblique rotation techniques as they would render more accurate and reproducible results when the factors are correlated.

Factor analysis produces two results: factor loading and factor scores. Factor loading represents the correlation of the original metrics with each identified factor, while factor scores are the values of each data point mapped in the factor space. We used the factor loading result to interpret the relations between metrics listed in Table \ref{Table:Metrics}. The factor scores were then used for the clustering analysis in the next step.

\subsubsection{Clustering}
\label{subsec:cluster}
After the activities (i.e. factors) were identified, we conducted a hierarchical clustering analysis based on the factor scores data to identify the prominent roles of OSS contributors. This method aims to construct a hierarchical structure of clusters; such structure provides more information about the dataset than unstructured clusters produced by flat clustering methods such as K-means. Furthermore, hierarchical methods do not require a predetermined number of clusters and most of them are deterministic. As such, this method supports the exploratory nature of our study.

Particularly, we used an agglomerative (or bottom-up) hierarchical clustering method. In general, agglomerative methods first treat each data point as a singleton cluster. Pairs of closest clusters are then successively merged until all clusters have been merged into a single one that contains all data. This process produces a hierarchy of clustering that can be visualized in a tree diagram named dendrogram. Cutting the dendrogram at a certain level creates a partition of disjoint clusters. This step is equivalent to grouping only the clusters with high similarity. Different strategies have been proposed for measuring the similarity between two clusters. Based on our initial experiment, we decided to use the ward’s method \cite{Murtagh2014}. This method produces clusters that are more compact and suitable for identifying and interpreting prominent roles. 

We used the silhouette value to measure the quality of clusters \cite{rousseeuw1987silhouettes}. It represents how similar one data point is to its own cluster compared to other clusters. To choose the optimal number of clusters, we considered the silhouette value while also referencing to the dendrogram produced by the ward hierarchical algorithm. 


\subsubsection{Interpreting activities and roles}
In order to identify the meaningful activities and roles represented in the factors and the clusters, we followed a qualitative process that involved the following steps. First, both authors independently examined the actual actions represented by the influential metrics for each factor and each wrote three to five keywords/phrases to describe their understanding of the factor. Then the authors discussed their notes and conducted an ``Affinity Diagraming'' study to group their keywords/phrases. Next, a phrase of higher-level abstraction was given to each group to describe the factor. Finally, the authors discussed and agreed on the phrase that described the biggest group in the affinity diagram of each factor as the activity it represented. We adopted a similar process in identifying the roles from the clustering analysis results.

\subsection{Analyzing Role Dynamics}
\label{subsec:Methods-Dynamics}

To identify patterns in the dynamics of changes in roles assumed by individual contributors, we first analyzed the frequency of changes among the roles with respect to all contributors. We then measured the role change intensity (RCI) for each contributor. A contributor's RCI was calculated by accumulating, over the 12 time periods, the quantity of role change between each two consecutive time periods; this quantity is measured using the Euclidean distance between cluster centroids of the two roles taken by the contributor in two consecutive time periods. To accommodate the large range of change intensity values and to ease comparison, we calculate RCI using a logarithmic scale. Therefore, the overall Role Change Intensity (RCI) for each contributor $i$ is:
\begin{equation} \label{eq:changeIntensity}
RoleChangeIntensity(i)=
\log_{10}{\sum_{t=2}^{12}dist(\mathbf{R}^i_t,\mathbf{R}^i_{t-1})}\\
\end{equation}
where $\mathbf{R}^i_t$ is the cluster centroid of the role assumed by contributor $i$ at time $t$ and $dist(\mathbf{A},\mathbf{B})=\sqrt{\sum_{n}(a_n-b_n)^2}$ represents the Euclidean distance between vectors $\mathbf{A}$ and $\mathbf{B}$. This measure provides an ordinal evaluation of the intensity of the contributors' role change. 
\section{Results}
\label{sec:ResultsAndDiscussion}
In the following sections, we first present our results on factor and clustering analyses. We then present findings on role dynamics.

\begin{table*}
\renewcommand{\arraystretch}{0.8}
\begin{threeparttable}

\centering
\caption{Factor loadings of the action metrics}
\label{Table:FactorAnalysis}
\small
\begin{tabular}{@{}l@{}|cccccc|cc@{}}
\toprule
\multicolumn{1}{c|}{\textbf{Metrics}} & \textbf{Factor1} & \textbf{Factor2} & \textbf{Factor3} & \textbf{Factor4} & \textbf{Factor5} & \textbf{Factor6} & \textbf{$h^2$} & \textbf{$u^2$} \\

\midrule
\# of commits made & 0.00& \cellcolor[HTML]{FFF2CC}1.03 & 0.02 & -0.07 & -0.04 & 0.05 & 0.997 & 0.003 \\
\# of line of code changed in the codebase & 0.00& \cellcolor[HTML]{FFF2CC}1.03 & 0.02 & -0.07 & -0.04 & 0.05 & 0.997 & 0.003 \\
\# of files worked on & 0.00& 0.26 & -0.07 & \cellcolor[HTML]{E2EFDA}0.35 & \cellcolor[HTML]{E2EFDA}0.45 & -0.17 & 0.575 & 0.425 \\
\# of pull requests (PRs) made & 0.06 & -0.04 & -0.11 & \cellcolor[HTML]{E2EFDA}0.45 & \cellcolor[HTML]{FFF2CC}0.70 & 0.05 & 0.876 & 0.124 \\
Avg. length of PR descriptions & 0.05 & -0.01 & -0.03 & -0.01 & 0.13 & 0.01 & 0.022 & 0.978 \\

\midrule
\# of issues reported & 0.30 & 0.04 & 0.18 & 0.11 & 0.11 & \cellcolor[HTML]{E2EFDA}0.35 & 0.643 & 0.358 \\
Avg. length of issue descriptions & -0.01 & 0.00& -0.03 & 0.00& 0.02 & 0.16 & 0.024 & 0.976 \\
\# of comments made in issue discussions & \cellcolor[HTML]{FFF2CC}0.52 & -0.01 & 0.30 & 0.24 & -0.13 & 0.16 & 0.900 & 0.100 \\
Avg. length of issue comments & -0.02 & 0.01 & 0.00& -0.01 & 0.00& 0.1 & 0.008 & 0.992 \\
\# of comments made in PR discussions & \cellcolor[HTML]{FFF2CC}0.72 & -0.01 & 0.09 & 0.03 & 0.25 & -0.12 & 0.820 & 0.180 \\
Avg. length of PR comments & 0.00& -0.01 & 0.04 & -0.02 & 0.09 & 0.03 & 0.013 & 0.987 \\

\midrule
\# of times being mentioned in issue comments & \cellcolor[HTML]{FFF2CC}0.78 & 0.01 & 0.1 & 0.01 & -0.02 & 0.14 & 0.840 & 0.160 \\
\# of times being mentioned in PR comments & \cellcolor[HTML]{FFF2CC}1.02 & 0.01 & -0.16 & -0.2 & 0.2 & -0.22 & 0.743 & 0.257 \\
\# of times referred other issues/PRs in issue comments & -0.12 & 0.02 & \cellcolor[HTML]{FFF2CC}1.06 & -0.07 & 0.02 & -0.15 & 0.808 & 0.192 \\
\# of times referred other issues/PRs in PR comments & 0.06 & 0.00& \cellcolor[HTML]{FFF2CC}0.59 & -0.05 & 0.21 & 0.00& 0.503 & 0.497 \\

\midrule
\# of times applied or removed labels on issues & 0.06 & 0.01 & \cellcolor[HTML]{FFF2CC}0.72 & 0.24 & -0.23 & -0.07 & 0.725 & 0.275 \\
\# of times applied or removed labels on PRs & \cellcolor[HTML]{FFF2CC}0.58 & 0.00& 0.13 & -0.03 & 0.15 & -0.07 & 0.516 & 0.485 \\
\# of times closed issues & 0.28 & -0.03 & 0.08 & \cellcolor[HTML]{FFF2CC}0.65 & -0.24 & 0.04 & 0.701 & 0.299 \\
\# of times closed pull requests & -0.30 & -0.08 & 0.07 & \cellcolor[HTML]{FFF2CC}0.95 & \cellcolor[HTML]{E2EFDA}0.33 & -0.06 & 0.857 & 0.144\\

\midrule
\multicolumn{1}{r|}{\textbf{Activity Name}}& \begin{tabular}[c]{@{}c@{}}Knwl. \\ Sharing\end{tabular} & \begin{tabular}[c]{@{}c@{}}Code \\ Contrib.\end{tabular} & \begin{tabular}[c]{@{}c@{}}Issue \\ Coord.\end{tabular} & \begin{tabular}[c]{@{}c@{}}Prog. \\ Ctrl.\end{tabular} & \begin{tabular}[c]{@{}c@{}}Code \\ Twking.\end{tabular} &
\begin{tabular}[c]{@{}c@{}}Issue \\ Rptg.\end{tabular}\\

\bottomrule
\end{tabular}
\begin{tablenotes}
\item Note 1: \textit{The $h^2$ column represents the estimated proportion of variance of the each metrics that are shared with other metrics and can explained by factors. The $u^2$ column (equals $1-h^2$) denotes the variance that are unique to the metric itself.}
\item Note 2: \textit{Yellow cells indicate that the loading is greater than 0.5; green cells indicate that the loading is between 0.3 and 0.5.}
\end{tablenotes}
\end{threeparttable}
\end{table*}

\subsection{Activity Extraction}
Based on the criterion introduced in Section \ref{subsubsec:factor}, we retained six factors that had eigenvalues greater than 1.0. These six factors explained 61\% of the data variance. The factor loading results are shown in Table \ref{Table:FactorAnalysis}. Based on the qualitative analysis described in Section \ref{subsec:Methods-Dynamics}, we explain the activities represented in these factors as follows:


\begin{table}[tbh]
\renewcommand{\arraystretch}{0.8}
\centering
\small
\caption{Correlation coefficient among factors}
\vspace{-5px}
\label{tab:factor_correlation}
\begin{tabular}{l|ccccc}
\toprule
& \begin{tabular}[c]{@{}c@{}}Knwl. \\ Sharing\end{tabular} & \begin{tabular}[c]{@{}c@{}}Code \\ Contri.\end{tabular} & \begin{tabular}[c]{@{}c@{}}Issue \\ Coord.\end{tabular} & \begin{tabular}[c]{@{}c@{}}Prog. \\ Ctrl.\end{tabular} & \begin{tabular}[c]{@{}c@{}}Code \\ Twking.\end{tabular} \\
\midrule
Code Contrib. & 0.18 & & & & \\ 
Issue Coord. & 0.70 & 0.09  & &  & \\ 
Progress Ctrl. & 0.61 & 0.35 & 0.56 & & \\ 
Code Twking. & 0.47 & 0.30 & 0.29 & 0.32  & \\ 
Issue Rptg. & 0.38 & -0.05 & 0.44 & 0.13  & 0.22 \\
\bottomrule
\end{tabular}
\vspace{-5px}
\end{table}

\renewcommand{\labelitemi}{$-$}
\begin{itemize}[leftmargin=*]
    \setlength\itemsep{2px}
    \item Factor 1 measures three types of actions: commenting, being mentioned in comments, and manipulating labels on PRs. The commenting actions may be associated with several purposes such as voicing opinions, providing suggestions, and asking or answering questions. But this factor is most heavily influenced by the number of times the contributor being mentioned; it also puts a heavy weight on label manipulation actions. These facts indicated that it mainly measures behaviors of providing information and knowledge about the project. We thus name this activity \textbf{Knowledge Sharing}.
    
    \item Factor 2 exclusively measures participants' contributions to the codebase. We name it \textbf{Code Contribution}.
    
    \item Factor 3 measures issue referring and label manipulating actions. We found that issue referring actions are usually associated with identifying duplicated issues or redirecting participants to move their discussion to other issues. At the same time, manipulating issue labels usually involve categorizing issues (e.g. into bugs or feature requests), identifying duplicated issues, and/or indicating stages in issue resolving progress (e.g. triaging, assigned). We thus name this activity \textbf{Issue Coordination}.
    
    \item Factor 4 is mostly associated with actions of closing issues or PRs. We name this activity \textbf{Progress Control}.
    
    \item Factor 5 is influenced by actions of making PRs and working on a large number of files. These indicate feature tweaking or bug fixing activities in which contributors make small changes on many files and file PRs for these changes to be included in the main repository. We thus name this activity \textbf{Code Tweaking}.
    
    \item Factor 6 is only influenced by the number of issues reported. We thus name it \textbf{Issue Reporting}.
\end{itemize}

The factor analysis result also demonstrated some correlations among the extracted activity dimensions (see Table \ref{tab:factor_correlation}). Particularly, Knowledge Sharing, Issue Coordination, and Progress Control exhibited high correlations (all pair-wise correlation coefficients $r>0.5$). Two other pairs of dimensions, Knowledge Sharing--Code Tweaking and Issue Reporting--Issue Coordination, also demonstrated moderate correlation ($r>0.4$). These results supported our hypothesis that factors influencing the contributors' actions in the OSS community are not independent.

\subsection{Roles Identification}
\begin{figure}[t]
	\centering
	\includegraphics[width=0.90\columnwidth]{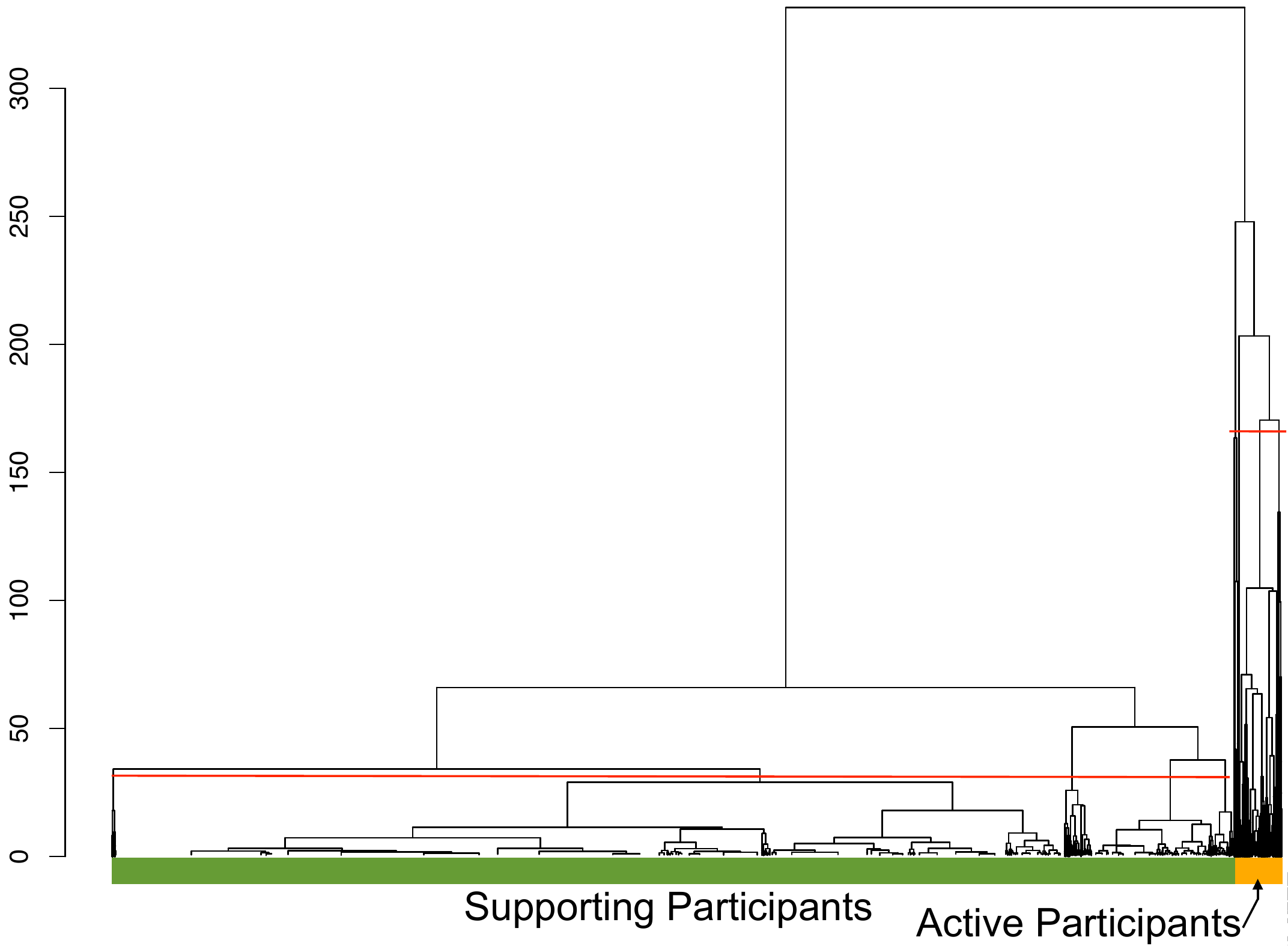}
    \caption{Hierarchical tree of clustering results. The red lines indicates the height cutoff values for the two groups.}
    \vspace{-15pt}
	\label{fig:HierarchicalTree}
\end{figure}

Figure \ref{fig:HierarchicalTree} shows the dendrogram of our hierarchical clustering results. We observed that there are two major groups of clusters that exhibited markedly different structures. The majority of the data points ($N=37,310$) fell into a cluster with a low dendrogram height, while some data points ($N=1,581$) represented a much higher height. In other words, the variance among sub-clusters in the first group was much smaller than that of the second group. This difference indicated that our data included two very distinct groups of users.

By examining the cluster centers and samples from each group, we found that the second group generates a much higher value in all factor dimensions when compared with the first group; in other words, contributors in this group are much more active in terms of all activities indicated by the factors. We thus consider the second group as comprised of \textbf{Active Contributors} in their communities, while the first group represents the \textbf{Supporting Contributors}. Because the sub-cluster distances within these two high-level groups are very different, we cut the two sub-trees at different heights when identifying the specific role clusters.

\begin{figure}[t]
	\centering
	\includegraphics[width=0.8\columnwidth]{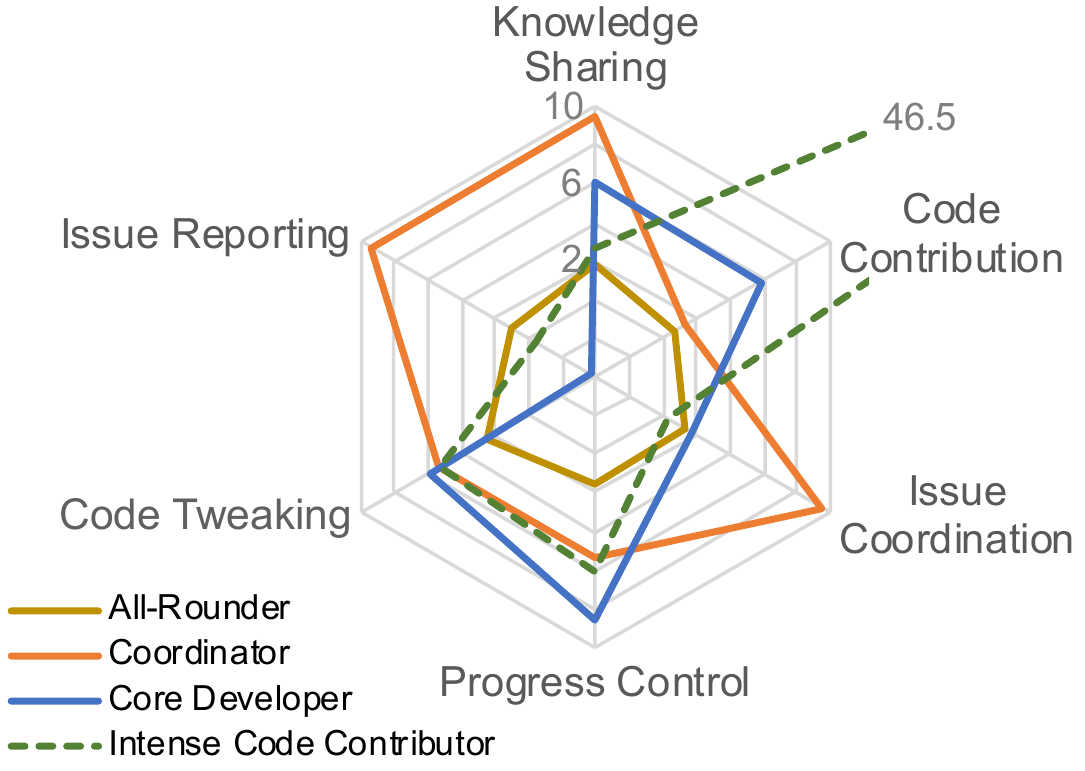}
    \caption{Activity space characteristics of roles in the Active Contributors group.}
	\label{fig:ActiveRoles}
\end{figure}

\begin{figure}[t]
	\centering
	\includegraphics[width=0.8\columnwidth]{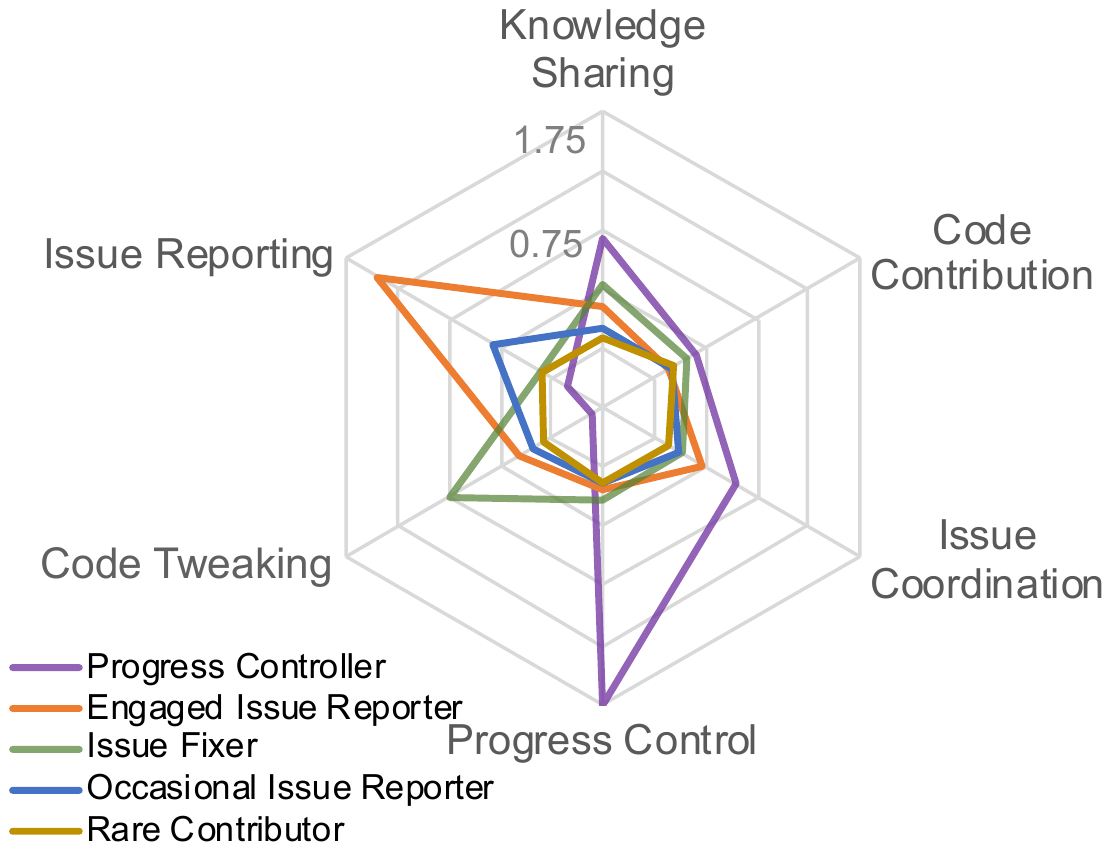}
    \caption{Activity space characteristics of roles in the Supporting Contributors group.}
	\label{fig:SupportingRoles}
\end{figure}

Based on the silhouettes measure and the dendrogram structure, we considered four clusters in the Active Contributors group and five clusters for Supporting Contributors; the red lines on Figure \ref{fig:HierarchicalTree} indicates the height cutoff values. Once the clusters were determined, we followed a qualitative process and named the clusters based on the activity space characteristics of each cluster centroid and analysis of actual actions performed by representative users in each cluster. The characteristics for those clusters are shown in Figure \ref{fig:ActiveRoles} and Figure \ref{fig:SupportingRoles} respectively. We discuss those clusters and our rationale for naming the roles as follows.

Among the \textit{Active Contributors}, \textbf{Intense Code Contributors} exerted an extremely high contribution to the codebase. Additionally, there was only a small number of contributors assumed this role. Their main focus seemed to be developing a certain functionality of the software within a short time period. \textbf{Coordinators} provided only a small amount of code contribution. Instead, they focused mainly on Knowledge Sharing, Issue Coordination, and Issue Reporting activities. They are usually the owner of the project or a core member of the community. \textbf{Core Developer} exerted very little contribution to Issue Reporting but performed actively in Code Contribution, Code Tweaking, Progress Control, and Knowledge Sharing. They seemed to focus mainly on development and knowledge sharing about the code. \textbf{All-Rounders} provided a medium level of contribution in all dimensions.

The \textit{Supporting Contributors} usually focused on only one or two activities. \textbf{Engaged Issue Reporters} and \textbf{Occasional Issue Reporters} both focused on Issue Reporting, but differed on the quantity of their contributions. \textbf{Progress Controllers} mostly engaged in the Progress Control activity, with some contribution to Knowledge Sharing and Issue Coordination; they almost never engaged in Code Tweaking contributions. An analysis of sample users of this group revealed that they are usually core members of the community and focused on activities such as code reviewing, quality control, and approving and merging PRs. \textbf{Issue Fixers} focused on making small tweaks to the code or fixing bugs. \textbf{Rare Contributors} only participated in a minuscule amount of activities.



\subsection{Role Dynamics}
Among all contributors included in our data set, most ($N=16,706$, $78.9\%$) only assumed the Rare Contributor role in certain periods of time during the past three years. There were also two contributors who engaged in their projects throughout the analyzed time periods with the same role (All-Rounder). We excluded them in our analysis. Within the rest of the contributors, there are 4,483 who only assumed roles in the Supporting role group throughout the 12 time periods. The rest ($N=479$) have assumed roles in the Active role group at least one time in the past three years. We focused on analyzing the role change dynamics of these two types OSS community contributors. When performing the analysis, we considered ``Absent'' (i.e. did not perform any contribution during a time period) as an additional role type.

\begin{figure}[ht]
	\centering
	\includegraphics[width=1\columnwidth]{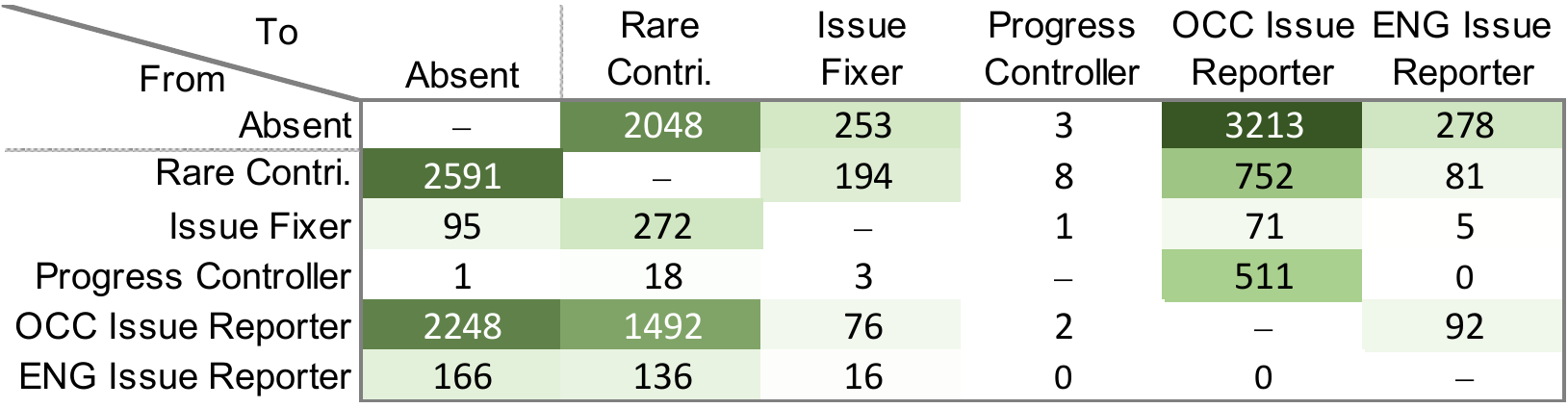}
    \caption{Role transition frequency heatmap for contributors who only assumed the \textit{Supporting} roles.}
	\label{fig:PersonalHeatmapTypeI}
\end{figure}

Figure \ref{fig:PersonalHeatmapTypeI} shows the heatmap of role transition frequency for contributors who only assumed the Supporting roles. Not surprisingly, the most frequency transition happened among Absent, Rare Contributor, and Occasional Issue Reporter roles. The transition frequency from Absent to the other roles also indicated ways people got involved to an OSS community: people rarely started as a Progress Controller; aside from occasional contributions, contributors usually began to engage in a project by assuming Issue Fixer and Engaged Issue Reporter roles. Interestingly, while any transition to the Progress Controller was rare, the transition from Progress Controller to Occasional Issue Reporter was frequent. This may illustrate a retiring path of community core members if they do not continue contributing as an Active Participant.

\begin{figure}[ht]
	\centering
	\includegraphics[width=1\columnwidth]{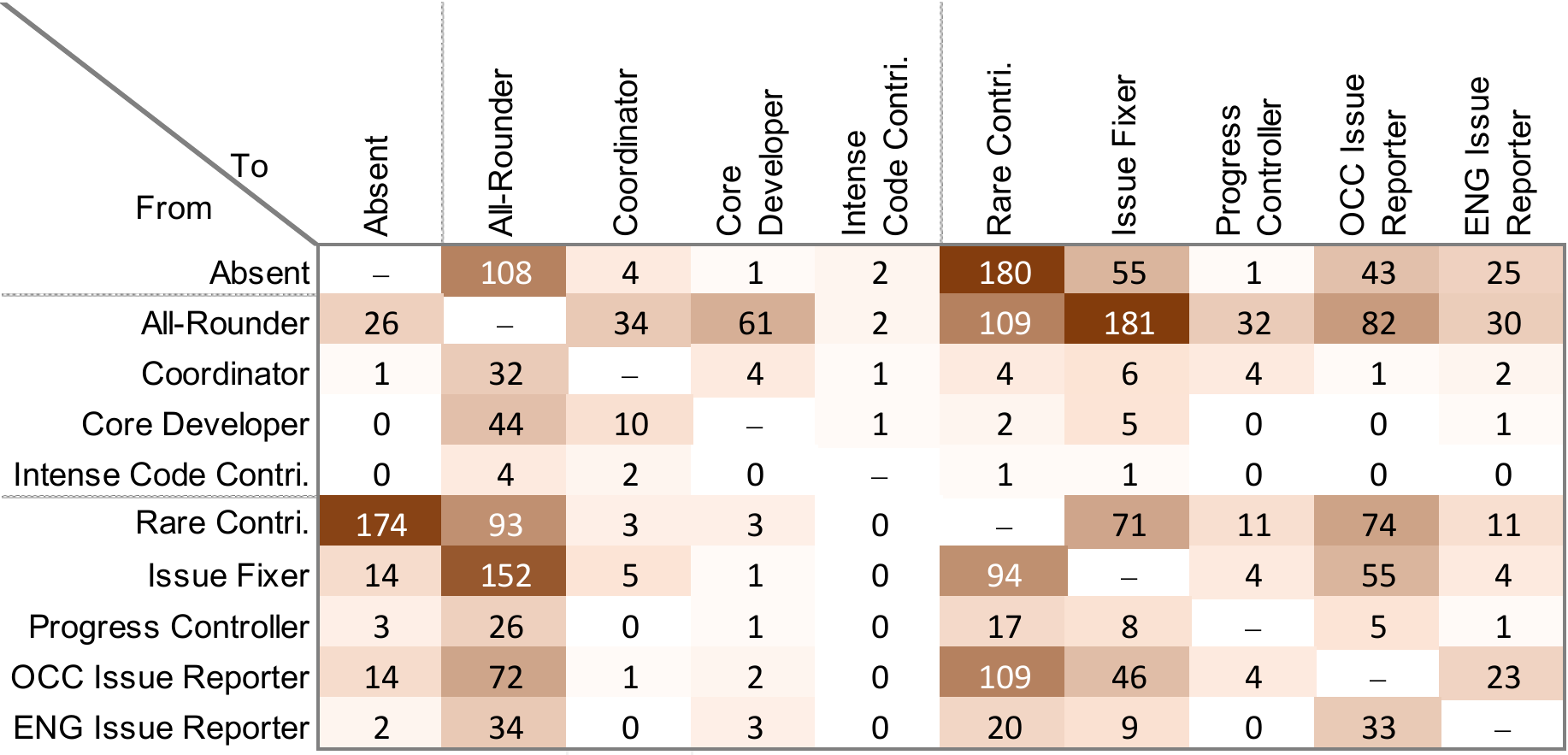}
    \caption{Role transition frequency heatmap for contributors who have assumed the \textit{Active} roles.}
	\label{fig:PersonalHeatmapTypeII}
\end{figure}

Figure \ref{fig:PersonalHeatmapTypeII} shows the heatmap of role transition frequency for contributors who have assumed the Active roles. Aside from transitions between Absent and Rare Contributors, the most frequent transitions happened between All-Rounder (Active role) and Issue Fixer (Supporting role). This type of transition may have represented a working style of a group of OSS contributors, who generally engaged in all aspects of the community but switch to focus on issue fixing when issues accumulate. Transitions among All-Rounder (Active role), Rare Contributor (Supporting role), and Occasional Issue Reporter (Supporting role) were also frequent, indicating many active community contributors may take breaks from their work.

\begin{figure}[h]
    \centering
    \small
	\includegraphics[width=0.8\columnwidth]{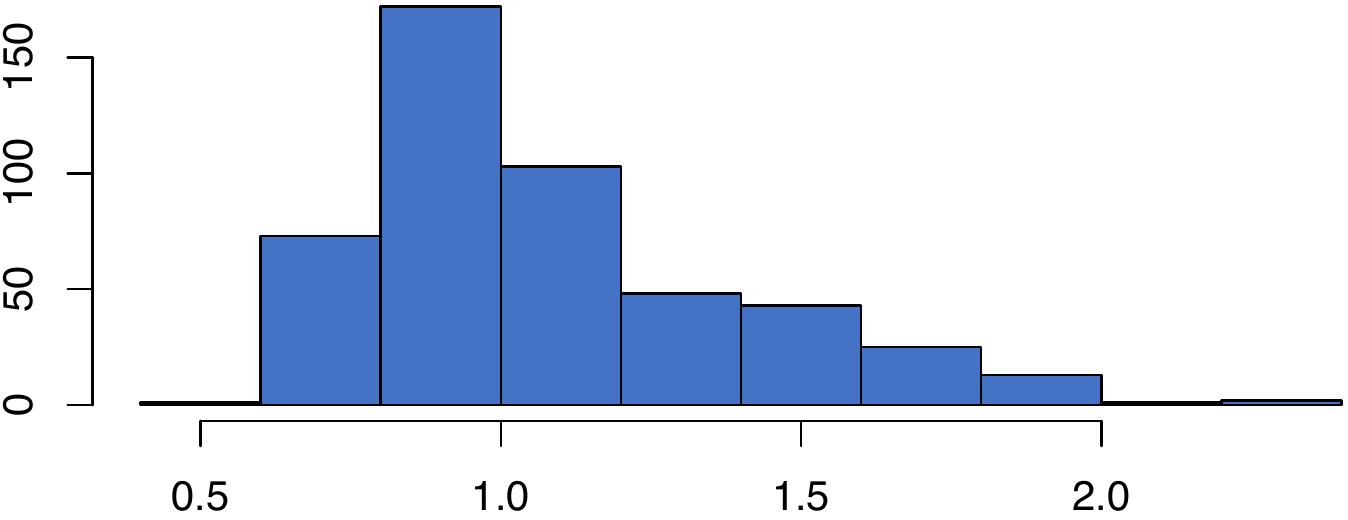}\\
    \caption{Histogram of RCI across contributors who have assumed the \textit{Active} roles.}
	\label{fig:RCIHisto}
	\vspace{-10px}
\end{figure}

We calculated a Role Change Intensity (RCI) score for each participant in each project using Equation \ref{eq:changeIntensity}. As the cluster distances among the Supporting roles are close to each other, the RCI scores for contributors who only assumed the Supporting roles are expected to be close to zero. As a result, we only focused on RCI scores for contributors who have ever assumed the Active roles. Figure \ref{fig:RCIHisto} shows the histogram of their RCI scores.
The results showed a right-skewed distribution with a median of 0.99 ($IQR=1.21-0.92$). Examining the scenario in which only one role change has occurred throughout the whole time periods, we found that the median RCI across all types of role changes is 1.11 ($IQR=1.26-0.35$). These results indicated that most contributors engaged in a medium-level RCI while some experienced role changes across a medium to high-level intensity.

\section{Discussion}
\label{sec:Implications}
Our work provides several implications for designing OSS tools to support role-based interactions. In this section, we first discuss these implications. We then consider the limitations of our current study and discuss directions for future work.

\subsection{Mediate OSS Activities Instead of Actions}
Current OSS tools usually involve features that focus on supporting low-level actions such as code committing, issue reporting, commenting, applying labels, etc. There is a lack of focus on mediating higher-level OSS activities. Through the factor analysis, we identified six activities that most clearly distinguished the roles of OSS contributors. Some of them, such as Code Contribution and Issue Reporting, only involved a small number of actions and were well supported in the current tools. However, others involved multiple actions that were currently supported in different, and sometimes isolated tool features. For example, the Issue Coordination activity involves actions to create links among the issues and the pull requests, as well as managing issue labels; on GitHub, there is no connection between the link creation (through commenting) and the issue labeling features.

Reflecting on the Activity Theory, tools need to serve as ``functional organs'' to help users achieve their goal-oriented activities \cite{Kaptelinin1996, Nardi1996}. We argue that the OSS tool designers need to consider the activities identified in this paper, which were aligned with the representative goals that OSS contributors hold when performing the corresponding actions. Particularly, they may explore connections surrounding the features that support the underlying actions for each activity to facilitate a smooth transition among the actions. Moreover, based on the correlations found among the activities, OSS tool designers may consider more sophisticated feature connections to support users move among activities.


\subsection{Support Role-Based Interaction}
Through the clustering analysis, we identified four Active roles and five Supporting roles of OSS contributors. While the literature has strongly advocated role-based interaction in software engineering tools \cite{Zhu2015}, the realization of such interaction is still immature in the OSS world. One reason for this gap is that there is currently little knowledge or guidance for the tool designers to have a clear conception as to what high-level activities and detailed actions each role takes. Our data-driven and activity-based roles addressed this limitation.

On one hand, the roles identified in this paper reflected some characteristics of the roles in the literature (e.g. the ``onion'' model \cite{Nakakoji2002}). For example, confirming the hypothesis posed in the ``onion'' model, our data indicated a small number of Active contributors who make a large amount of contributions and a vast number of Supporting contributors. On the other hand, however, our roles provided a non-simplistic trace to the main activities each role focuses on. For example, our results showed that the \textit{Progress Controllers} do not only perform the Progress Control activity but they were usually also involved in Knowledge Sharing and Issue Coordination; the \textit{Engaged Issue Reporters} usually also perform the Issue Coordination activity. As a result, these roles paint a more comprehensive picture about activities and responsibilities of OSS contributors. The OSS tool designers can use this information to better satisfy the goals and needs of OSS contributors in role-based interaction design. Particularly, they can use the activities and the actions underlying each role as a design guideline.

\subsection{Support Role Change: Onboarding and Retiring}
There is limited discussion about OSS tools that support role change in the literature. However, our data indicated that role change in OSS communities is both frequent and somewhat intense. As a result, techniques and tools that facilitate a smooth change of roles can be useful for OSS contributors.

While our results have indicated a complex role change model, onboarding and retiring are among the most important types of role change for OSS communities. Our data confirmed a common impression that OSS contributors usually get involved in a project through issue reporting and fixing. We also identified that a common retiring path of Active contributors is also though issue-related activities. These findings indicated a central role of the Issue Management Systems in the onboarding and retiring processes. Tool designers may consider including features in the Issue Management Systems to support new contributors to be better engaged in the community culture and acquire the necessary knowledge and skill. They may also enhance the Issue Management Systems to help retiring members transfer knowledge and tasks.

\subsection{Limitations and Future Work}
Although diverse, the contribution metrics used in our study are based only on quantity, rather than quality, of the actions taken by OSS contributors. Our research can thus be extended with studies focused on extracting qualitative measures of contribution. Additionally, while the 29 OSS projects analyzed in this work were carefully selected to cover a wide variety in terms of application domains, programming language, community size, and code base size, future work that validates our findings in a larger amount of OSS projects and communities can be useful. Moreover, we focused on investigating OSS contributors' roles within a project. However, many contemporary OSS communities were structured around a group of projects (i.e. a project ecosystem). Exploring how our model and method can generalize to such higher-scaled OSS communities is an interesting future work.
\section{Conclusion}
\label{sec:Conclusion}
In this study, we adopted a data-driven approach to understanding the diverse roles and their dynamics in OSS communities. From an analysis of 29 OSS projects, we extracted six activities that determined four Active roles and five Supporting roles. This approach allowed us to provide rich information, grounded in the data, about the actions and activities performed by each role. Through the lens of the Activity Theory, such information rendered useful design guidelines for role-based OSS tools. We argue that such methodology and the generated information are crucial to understanding and supporting the collaboration among diverse OSS contributors.

\bibliographystyle{abbrv}

\end{document}